\documentclass[12pt,notoc]{JHEP3}

\usepackage{amsmath,amssymb,euscript,array,cite,mathrsfs}

\def\oldf{f}

\newcommand{\startappendix}{
\setcounter{section}{0}
\renewcommand{\thesection}{\Alph{section}}}
\newcommand{\Appendix}[1]{
\refstepcounter{section}
\begin{flushleft}
{\large\bf Appendix \thesection: #1}
\end{flushleft}}
\usepackage{epsfig}

\newcommand{\cosec}{\operatorname{cosec}}

\newcommand{\Tr}{\operatorname{Tr}}

\def\Bv{{\boldsymbol v}}

\def\Bvarpi{{\boldsymbol \varpi}}

\def\B0{{\boldsymbol 0}}
\def\BF{{\boldsymbol F}}

\def\BOmega{{\boldsymbol\Omega}}
\def\Bvarpi{{\boldsymbol\varpi}}

\def\CP{{\mathbb C}P}

\def\Tr{{\rm Tr}}

\def\R{{\mathbb R}}

\newcommand{\Be}{\boldsymbol{e}}

\def\Dbarslash{\,\,{\raise.15ex\hbox{/}\mkern-12mu {\bar D}}}
\def\Dslash{\,\,{\raise.15ex\hbox{/}\mkern-12mu D}}
\def\delslash{\,\,{\raise.15ex\hbox{/}\mkern-9mu \partial}}
\def\delbarslash{\,\,{\raise.15ex\hbox{/}\mkern-9mu {\bar\partial}}}

\def\LAG{\mathscr{L}}

\newcommand{\MAT}[1]{\begin{pmatrix} #1\end{pmatrix}}

\newcommand{\EQ}[1]{\begin{equation}\begin{split} #1
\end{split}\end{equation}}

\newcommand{\SP}[1]{\begin{equation}\begin{split} #1
\end{split}\end{equation}}

\title{The Semi-Classical Spectrum of Solitons and Giant Magnons}
\author{Timothy J. Hollowood\\
Department of Physics,\\ University of Wales Swansea,\\
Swansea, SA2 8PP, UK.\\
E-mail: \email{t.hollowood@swansea.ac.uk}}
\author{and J.~Luis Miramontes\\
Departamento de F\'\i sica de Part\'\i culas and IGFAE,\\
Universidad
de Santiago de Compostela\\ 15782 Santiago de Compostela, Spain\\
E-mail: \email{jluis.miramontes@usc.es}}

\abstract{In this note, we summarize recent progress in constructing and then semi-classically quantizing solitons, or non-abelian Q-balls, in the 
symmetric space sine-Gordon theories. We then consider the images of these solitons 
in the related constrained sigma model, which are the dyonic giant magnons on the string theory world-sheet.
Focussing on the case of the symmetric space $S^5$, we perform a semi-classical quantization of the solitons and magnons and show that both lead to Chern-Simons quantum mechanics on the internal moduli space which is a real Grassmannian $SO(4)/SO(2)\times SO(2)$ but---importantly---with a different coupling constant. Quantizing this system shows that both the Q-balls and 
magnons come in a tower of states transforming in symmetric representations of the $SO(4)$ symmetry group; however, the former come in a finite tower whereas the latter come in the well-known infinite tower of dyonic giant magnons.}

\begin{document}

\section{Introduction}
\label{Intro}

There has been a great deal of progress in understanding the world-sheet theory of the string moving on some particular curved spacetimes like those involved in the basic AdS/CFT correspondence (see for example the series of review articles \cite{Beisert:2010jr}). The reason is that, under special circumstances, when the spacetime is of the form $\R_t\times F/G$, with $F/G$ a symmetric space like $S^n$ or $\CP^n$,  the world-sheet theory is a non-relativistic integrable system. The excitations of the string are known as giant magnons~\cite{Hofman:2006xt,Dorey:2006dq,Chen:2006gea,Gaiotto:2008cg,Grignani:2008is,Abbott:2009um,Hollowood:2009sc}, which are soliton-like solutions on the string world-sheet, and in many cases the exact factorizable S-matrix is already known~\cite{Staudacher:2004tk,Beisert:2005tm,Beisert:2006qh,Arutyunov:2006yd,Ahn:2008aa}. More precisely the giant magnons are kink solutions that correspond to open strings and have to be put together to make closed string configurations.
It was noted a long time ago that the gauge-fixed theory of the string on such spacetimes is classically equivalent to a relativistic $1+1$-dimensional integrable QFT~\cite{Tseytlin:2003ii,Mikhailov:2005qv,Mikhailov:2005sy} known as a symmetric space sine-Gordon (SSSG) theory. These theories arise as the result of imposing the Pohlmeyer reduction on a sigma model with target space a symmetric space $F/G$~\cite{Pohlmeyer:1975nb}, and their Lagrangian formulation was originally proposed in~\cite{Bakas:1995bm} (for a recent review see~\cite{Miramontes:2008wt} and references therein). Integrable systems typically have more than one compatible symplectic structures, and it is known that at the classical level
the string sigma model and the SSSG theories have different symplectic structures \cite{Mikhailov:2006uc}. However, it has been suggested that the SSSG theory which is classically equivalent to superstrings on $AdS_5\times S^5$ could also be equivalent at the quantum level~\cite{Grigoriev:2007bu,Mikhailov:2007xr} (see also~\cite{Grigoriev:2008jq,Roiban:2009vh,Hoare:2009rq,Hoare:2009fs,Iwashita:2010tg,Hoare:2010fb}).

In this note we summarize some recent progress in understanding the SSSG theories and their relation to the string sigma models~\cite{Hollowood:2010dt}. We shall focus on the particular example of the SSSG theory associated to the symmetric space
\EQ{
S^5=F/G=\frac{SO(6)}{SO(5)}\ ,
}
which is relevant to the bosonic sector of the string moving on $AdS_5\times S^5$. We shall ignore the fermionic sector of these theories in the present letter, however, on the string side it is only the full theory with all the fermionic fields present which is expected to be a finite theory. For present purposes, where we only work at the level of semi-classical effects, we can ignore the issue of fully quantum effects and consequently ignore the fermions.\footnote{The symmetric space sine-Gordon theories with fermions has been considered in \cite{Schmidtt:2010bi}, and the solitons are constructed and investigated in \cite{Hollowood:2011fq}.}
The main paper \cite{Hollowood:2010dt} describes the generalization to any Symmetric Space of Type~I in Cartan's classification. The approach adopted in this work is to use the algebraic formalism of the symmetric space, and to this end we work with group- or algebra-valued fields 
in $F=SO(6)$, or $\mathfrak{so}(6)$. The subgroup $G=SO(5)$, or algebra $\mathfrak{so}(5)$, is defined as the subgroup/algebra fixed by the involution 
\EQ{
\sigma_-(f)=\theta f\theta\ ,\qquad\theta=\text{diag}(-1,1,1,1,1,1)\ ,
}
and in the following we will need the triplet of groups $SO(4)\subset SO(5)\subset SO(6)$ embedded as follows
\EQ{
SO(4)=\left(\begin{array}{cccccc} 1 & 0 & 0 &0&0&0\\ 0& 1 & 0&0&0&0\\ 0 & 0 & * &*&*&*\\ 0 & 0 & * &*&*&*\\ 0 & 0 & * &*&*&*\\ 0 & 0 & * &*&*&*\end{array}\right)\ ,\qquad
SO(5)=\left(\begin{array}{cccccc} 1 & 0 & 0 &0&0&0\\ 0& * & *&*&*&*\\ 0 & * & * &*&*&*\\ 0 & * & * &*&*&*\\ 0 & * & * &*&*&*\\ 0 & * & * &*&*&*\end{array}\right)\ .
\label{hgg}
}

\noindent
{\bf The Symmetric Space Sine-Gordon Theories}

For the symmetric space $S^5$, the SSSG theory is a gauged WZW theory for $SO(5)/SO(4)$ (where the anomaly free vector subgroup $H=SO(4)$ with  $\gamma\to U\gamma U^{-1}$ is gauged) deformed with a kind of mass (or potential) term. The action takes the form
\EQ{
S=S_\text{gWZW}[\gamma,A_\mu] +S_\text{bt}[\phi,A_\mu] -\frac k{2\pi}\int d^2x\,\Tr\left(\Lambda
\gamma^{-1}\Lambda\gamma-\Lambda^2\right)\ ,
\label{ala}
}
and is invariant under the gauge transformations
\SP{
\gamma\to U\gamma U^{-1}\,,\qquad A_\mu \to U\big(A_\mu +\partial_\mu\big)U^{-1}\,, \qquad U\in H\,.
\label{gaugeT}
}
The mass term involves the element $\Lambda$ of the Lie algebra $\mathfrak{so}(6)$, which up to conjugation takes the form
\EQ{
\Lambda=m\left(\begin{array}{cccccc} 0 & -1 & 0 &0&0&0\\ 1& 0 & 0&0&0&0\\ 0 & 0 & 0&0&0&0\\ 0 & 0 & 0 &0&0&0\\ 0 & 0 & 0 &0&0&0\\ 0 & 0 & 0 &0&0&0\end{array}\right)\ .
\label{lam}
}
Note that $H=SO(4)$, defined as in \eqref{hgg}, is the stability group of $\Lambda$.
Here, $S_\text{gWZW}[\gamma,A_\mu]$ is the usual gauged WZW action for $G/H$ with level $k$,
\EQ{
S_\text{gWZW}[\gamma,A_\mu]&=-\frac k{4\pi}\int d^2x\,\Tr\,\Big[
\gamma^{-1}\partial_+\gamma\,\gamma^{-1}\partial_-\gamma+2A_+\partial_-\gamma\gamma^{-1}\\ &~~~~~~~~~
-2A_-\gamma^{-1}\partial_+\gamma-2\gamma^{-1}A_+\gamma A_-+2A_+A_-\Big]
\\ &~~~~~~~~~+\frac{k}{24\pi}\int d^3x\,\epsilon^{abc}\Tr\,\Big[\gamma^{-1}\partial_a\gamma\,
\gamma^{-1}\partial_b\gamma\,\gamma^{-1}\partial_c\gamma\Big]\ ,
\label{gWZW}
}
and
\EQ{
S_\text{bt}[\phi,A_\mu]=-\frac{k}{4\pi} \int d^2x\> \epsilon^{\mu\nu}\partial_\mu \Tr\big(A_\nu \phi \big)\ ,\qquad \gamma = e^\phi\ .
\label{Topol}
}
is a (total derivative) boundary term which does not contribute to the equations of motion.

Below we note some features of this theory:

(i) It can be thought of as a deformation of the CFT given by the gauged WZW model by the particular relevant operator corresponding to the mass term~\cite{Bakas:1993xh,Bakas:1995bm,CastroAlvaredo:2000kq}.

(ii) Classically, the vacuum is degenerate and, for $A_\mu=0$, 
the potential term in~\eqref{ala} has a space of minima given by constant 
group elements $\gamma\in H=SO(4)$. Hence, classically at least, there could be a Higgs effect since $\gamma(x=\pm\infty)\in SO(4)$ and the $SO(4)$ global gauge symmetry is generically spontaneously broken. However, there are solitons in the theory in the form of kinks which have a gapless spectrum whose existence means that in the functional integral one should integrate over the boundary values $\gamma(x=\pm\infty)$. This has the effect of restoring the $SO(4)$ global symmetry associated to gauge transformations. In other words, the kinks carry $SO(4)$ global charge and so are dyonic objects; namely, non-abelian Q-balls \cite{Safian:1987pr}.
At the quantum level the continuous spectrum of kinks becomes quantized and the Q-ball states transform in non-trivial representations of $SO(4)$. In particular, in these theories the gauge symmetry is not confined and physical states carry ``colour".

(iii) The SSSG theories are integrable. This can be seen by 
writing the equations-of-motion of the SSSG equations in Lax form, that is as a zero curvature condition for a connection that depends on an auxiliary
complex spectral parameter $z$:
\EQ{
{\cal L}_\mu=\partial_\mu+{\cal A}_\mu(x;z)\ ,\qquad[{\cal L}_\mu(z),{\cal L}_\nu(z)]=0\ ,
\label{zcc}
}
where 
\SP{
{\cal L}_+(z)&= \partial_++\gamma^{-1}\partial_+\gamma+\gamma^{-1}A_+\gamma-z
\Lambda
\ ,\\[5pt]
{\cal L}_-(z)&= \partial_-+A_--z^{-1}\gamma^{-1}\Lambda\gamma\ .
}
The existence of the Lax connection implies integrability in a way which is completely standard.
In addition, the equation-of-motion of the gauge field yields the additional constraints
\SP{
\Big(\gamma^{\mp1}\partial_\pm\gamma^{\pm1}
+\gamma^{\mp1}A_\pm\gamma^{\pm1}\Big)^\perp=A_\pm\ ,
\label{gco2}
}
where $\perp$ is a projection onto the Lie algebra of H, which in this case is $\mathfrak{so}(4)\subset\mathfrak{so}(6)$.
Since the equations-of-motion imply that $A_\mu$ is a flat connection, one can fix the gauge on-shell by choosing $A_\mu=0$ (fixing the gauge off-shell is described in \cite{Hoare:2009fs,Hoare:2010fb}).

(iv) Since the fields do not fall-off at $x=\pm\infty$, the WZ term requires careful treatment. In particular, it cannot strictly speaking be defined as an integral over a three-dimensional space with the two-dimensional spacetime as a boundary, and its definition in~\eqref{gWZW} should be taken to be schematic. One way to unambiguously define the action is, as in~\cite{Hoare:2009fs}, to use the condition of gauge invariance to pin down the expansion of the WZ term in terms of $\phi$, with $\gamma=e^\phi$. This prescription requires to supplement the action with the boundary term~\eqref{Topol} in order to make the leading order contribution gauge invariant. To spell this out, notice that
\EQ{
&S_\text{gWZW}=-\frac k{2\pi}\int d^2x\,\Tr\,\Big(
A_+\partial_-\phi -A_-\partial_+\phi
+[A_+, A_-]\phi + \cdots\Big)
} while
\EQ{
S_\text{gWZW}+S_\text{bt}=-\frac{k}{2\pi} \int d^2x\,\Tr\,\Big(
[\partial_+ + A_+, \partial_- +A_-]\phi + \cdots\Big)\,,
}
which shows that $S_\text{gWZW}+S_\text{bt}$ is indeed invariant under~\eqref{gaugeT} at leading order. The explicit expression for the expansion of the full action~\eqref{ala} in powers of $\phi$ can be found in~\cite{Hoare:2009fs}.

(v) The constraints~\eqref{gco2} have the interpretation of the vanishing, on-shell, of what is naively the Noether current corresponding to global gauge transformations. In fact this is just an example of the theorem of Hilbert and Noether that the current associated to a local symmetry vanishes on-shell, but crucially up to a topological contribution which, in our case, is fixed by the boundary term~\eqref{Topol}.\footnote{For a discussion of these issues in a modern context see, for example, \cite{Julia:1998ys,Silva:1998ii,Julia:2000er} and references therein.}  The full expression for the current is
\SP{
{\cal J}_\pm=\frac{k}{4\pi} \Big(-(\gamma^{\mp1}\partial_\pm\gamma^{\pm1}+\gamma^{\mp1}A_\pm\gamma^{\pm1} -A_\pm)^\perp\pm\partial_\pm\phi^\perp\Big)\ ,
\label{ncor}
}
which exhibits that the Noether charge emerges as a kink charge,
\EQ{
{\cal Q}=\int dx\, {\cal J}^0= \frac{k}{4\pi}\int dx\,\partial_1\phi^\perp=
\frac{k}{4\pi}\big(\phi^\perp(\infty)-\phi^\perp(-\infty)\big)\ .
\label{chgs1}
}

\noindent
{\bf The String Sigma Model}

Once the string world-sheet theory for strings on ${\mathbb R}_t\times S^5$ is suitably gauge fixed, what remains is a sigma model with a target space $S^5\simeq SO(6)/SO(5)$ that can be 
formulated in terms of a group-valued field $f\in SO(6)$ and a gauge field $B_\mu\in\mathfrak{so}(5)$ with a gauge symmetry
\EQ{
\oldf \rightarrow \oldf  U^{-1}\>, \qquad
B_\mu \rightarrow U(B_\mu +\partial_\mu)U^{-1}\>,\qquad U\in SO(5)\>.
\label{GaugeTrans}
}
The action takes the form
\EQ{
S[f,B_\mu]= -\frac{\sqrt\lambda}{4\pi}\int d^2x\, \mathop{\rm Tr} \bigl(J_\mu J^\mu\bigr)\>,
\label{LagSM}
}
where the current $J_\mu = \oldf ^{-1}\partial_\mu \oldf -B_\mu
\rightarrow U J_\mu U^{-1}$ is covariant under gauge transformations, and $\lambda$ is the 't~Hooft coupling. The string sigma model also involves imposing the Virasoro constraints which, up to conjugation, take the form
\EQ{
J_+=-\Lambda\ ,\qquad J_-=-\gamma^{-1}\Lambda\gamma\ ,
\label{virc}
}
where the field $\gamma$ takes values in $G=SO(5)$.
The Virasoro constraints can be written as the auxiliary linear system \cite{Miramontes:2008wt}
\EQ{
\Big(\partial_++B_+-\Lambda\Big)f^{-1}=\Big(\partial_-+B_--\gamma^{-1}\Lambda\gamma\Big)f^{-1}=0\ .
\label{sml}
}

This sigma model has the vacuum solution
\EQ{
f_0=\exp(-2t\Lambda)\,.
\label{vacu}
}
Physically, it corresponds to a point-like string orbiting around the great circle in $S^5$ picked out by the element $\Lambda$. With our choice of $\Lambda$ in \eqref{lam} the motion is in the $(1,2)$ plane.
The sigma model has soliton solutions which, in their most general form, are known as dyonic giant magnons~~\cite{Hofman:2006xt,Dorey:2006dq,Chen:2006gea}. These solutions are also kinks because they describe open strings whose endpoints at $x=\pm\infty$ are at distinct points on $S^5$.

Notice that contrary to the SSSG theory the gauge symmetry is not realized in the spectrum, rather it is confined and one can use an equivalent manifestly gauge invariant formalism as in \cite{Hollowood:2009tw} by considering the gauge invariant field ${\cal F}=\theta f\theta f^{-1}$. As mentioned above, we shall ignore quantum effects and the running of the coupling in the sigma model.

\noindent
{\bf Relation between the string sigma model and the SSSG theory}

The sigma model and the SSSG theory are related via their equations-of-motion. In order to find the relation, notice that the Lax formulation of the SSSG equations \eqref{zcc} are the consistency conditions for the linear system
\EQ{
{\cal L}_\mu(z) \Upsilon(z)=0\ .
\label{LinProb}
}
It is useful to express
\EQ{
\Upsilon(z)=\chi(z)\Upsilon_0(z)\ ,
}
where, in the on-shell gauge $A_\mu=0$,
\EQ{
\Upsilon_0(x;z)=\exp\big[(zx^++z^{-1}x^-)\Lambda\big]
\label{dvs}
}
is the vacuum solution of the linear problem corresponding to $\gamma_0=1$. Then,
\EQ{
\gamma=\chi(0)^{-1}\ .
}

Comparing this linear system with \eqref{sml}, it follows that 
$\Lambda$ and $\gamma$ are identified with the same quantities in the SSSG equations, and that the gauge fields in the sigma model are identified via
\EQ{
B_+=\gamma^{-1}\partial_+\gamma+\gamma^{-1}A_+\gamma\ ,\qquad B_-=A_-\ .
\label{Match}
}
The sigma model equations-of-motion then imply that $\gamma$ satisfies the SSSG equations-of-motion.
In addition, the group field of the sigma model is simply
\EQ{
f=\Upsilon(1)^{-1}=\Upsilon_0(1)^{-1}\chi(1)^{-1}\ .
\label{xll}
}
The fact that the expression for $f$ involves fixing the spectral parameter manifests the fact that the sigma model is not relativistically invariant. The reason is that the spectral parameter $z$ transforms as $z\to e^\vartheta z$ under a Lorentz transformation, and so setting it to 1 breaks Lorentz symmetry explicitly. In \eqref{xll}, notice that $f_0=\Upsilon_0(1)^{-1}$ is the vacuum solution \eqref{vacu}.

The implication is that there is one underlying integrable system which is expressed in two different ways, one as a relativistic QFT and the other as the non-relativistic theory on the string world sheet. 

\section{The Q-ball Kinks and Dyonic Giant Magnons}
\label{Dressing}

A soliton solution of the integrable system is a Q-ball kink of the SSSG equations or a dyonic giant magnon of the string sigma model.\footnote{We shall reserve the term ``soliton" for the generic solution of the integrable system, and use ``Q-ball kink" and ``dyonic giant magnon" for its expression in the SSSG theory, via $\gamma$, and the sigma model, via $f$, respectively.} The explicit solutions are characterized by the dressing factor $\chi(z)$ having a small number of poles on the complex $z$ plane, four in the present case~\cite{Hollowood:2009sc}. In general, the pattern of the poles and the relation between the residues is determined by the particular symmetric space:
\EQ{
\chi(z)=1+\frac{\BF_j(\Gamma^{-1})_{ji}\BF_i^\dagger}{z-\xi_i}\ .
}
For $S^5$, $\BF_i$ are four complex 6-vectors given by
\EQ{
\BF_i=\Upsilon_0(\xi_i^*)\Bvarpi_i\ ,\qquad i=1,\ldots,4\ ,
}
where $\Bvarpi_i$ are four constant 6-vectors given by
\EQ{
\Bvarpi_i=\{\Bvarpi,\Bvarpi^*,\theta\Bvarpi,\theta\Bvarpi^*\},
\label{vwr}
}
with the constraint
\EQ{
\Bvarpi\cdot\Bvarpi=0\ .
}
Fixing various redundancies, for a single soliton the constant 6-vector $\Bvarpi$ can be written as
\EQ{
\Bvarpi=\Bv+\BOmega\ .
}
Here, $\Bv$ is one of the non-null eigenvectors of $\Lambda$. Since it does not matter which one is chosen, we take
\EQ{
\Bv=\frac1{\sqrt{2}}(1,-i,0,0,0,0)\ .
}
The other vector $\BOmega$ is a unit complex vector in the 4-dimensional subspace picked out by the subalgebra $\mathfrak{so}(4)\subset\mathfrak{so}(6)$, so that $\BOmega$ is a null eigenvector of $\Lambda$ with the additional constraint $\BOmega\cdot\BOmega=0$. The positions of the four poles are 
\EQ{
\xi_i=\{\xi,\xi^*,-\xi,-\xi^*\}
\label{Poles}
}
and, finally, the matrix $\Gamma$ reads
\EQ{
\Gamma_{ij}=\frac{\BF_i^*\cdot \BF_j}{\xi_i-\xi^*_j}\ .
}

The data $(\BOmega,\xi)$  are the parameters associated to
the solution. The complex variable $\xi$ determines the energy and momentum of the soliton whereas $\BOmega$ is a genuine collective coordinate which labels solutions of the same energy and momentum. Writing $\xi =e^{-\vartheta -iq}$, the $t$ and $x$ dependence of the the soliton follow from
\EQ{
\BF_1=  \Upsilon_0(\xi^\ast) \Bvarpi=\exp\left[2imt' \cos q - 2mx' \sin q\right]\Bv+\BOmega\ ,
\label{Fvector}
}
where we have introduced the boosted cordinates
\EQ{
t' = t\cosh\vartheta -x\sinh\vartheta\ ,\qquad
x' = x\cosh\vartheta -t\sinh\vartheta
}
which identify $\vartheta$ as the rapidity.

The soliton carries a non-trivial moduli space corresponding to 
the ``polarization" vector $\BOmega$. We can write $\BOmega=\frac1{\sqrt2}(\BOmega_1+i\BOmega_2)$, where $\BOmega_i$ are two real orthonormal vectors.
The moduli space is swept out by the action of the global $SO(4)$ part of the gauge symmetry group identified with a particular (co-)adjoint of the form
$Uh_\BOmega U^{-1}$, for 
\EQ{
h_\BOmega=i(\BOmega\BOmega^\dagger-\BOmega^*\BOmega^t)=\BOmega_1\BOmega_2^t-\BOmega_2\BOmega_1^t \in \mathfrak{so}(4)\ .
\label{defh}
}
Up to conjugation, we may take $\BOmega=\frac1{\sqrt2}(0,0,1,-i,0,0)$,  which gives an orbit of the form 
\EQ{
U\MAT{0&0&0&0&0&0\\ 0&0&0&0&0&0\\ 0&0&0&-1&0&0\\ 0&0&1&0&0&0\\ 0&0&0&0&0&0\\ 0&0&0&0&0&0}U^{-1}\ ,
}
where $U\in SO(4)$ as in \eqref{hgg}.
This identifies the orbit as the real Grassmannian
\EQ{
\mathfrak M=\frac{SO(4)}{SO(2)\times SO(2)}\simeq S^2\times S^2\ .
\label{fgt}
}

On the SSSG side, $q$ determines the mass and charges of the Q-ball kink. Physically, distinct solutions are given by restricting $q\in(0,\frac\pi2)$ and they have mass 
\EQ{
M=\frac{4km}\pi\sin q\ .
}
Their kink charge is 
\EQ{
\gamma^{-1}(x=\infty)\gamma(x=-\infty)=\exp\left[-4qh_\BOmega\right]\ .
}
Assuming that $\gamma(-\infty)$ and $\gamma(+\infty)$ commute, which is true for these configurations, the kink charge corresponds to
\EQ{
{\cal Q}=\frac{k}{4\pi}\big(\phi^\perp(\infty)-\phi^\perp(-\infty)\big) =\frac{k q}{\pi}\,h_\BOmega
}
in~\eqref{chgs1}. Therefore, since this is the Noether charge under global gauge transformations, these solutions can actually be understood both as kinks and as non-abelian $Q$-balls~\cite{Safian:1987pr}. 

On the sigma model side, the dyonic giant magnon is characterized by its charge under the global symmetry corresponding to left multiplication $f\to Uf$, $U\in F$, relative to the vacuum solution. The conserved current for left multiplication is $D_\mu f\,f^{-1}$, and so
\EQ{
\Delta{\cal Q}_L=\int_{-\infty}^\infty dx\,\big(D_0f\,f^{-1}-D_0f_0\,f_0^{-1}\big)\ ,
}
where the vacuum solution is
\EQ{
f_0=\exp(-2t\Lambda)\ .
}
The calculation of this charge is described in Appendix~\ref{AppA}. It consists of two distinct contributions:
\EQ{
\Delta{\cal Q}_L = J_1m^{-1}\Lambda -J_2h_{\BOmega}\ ,
\label{NewCharge2}
}
with
\EQ{
J_1=\frac{r^2+1}r\Big|\sin\frac p2\Big|\ ,\qquad J_2=\frac{r^2-1}r\Big|\sin\frac p2\Big|\ ,
}
where the parameter $re^{ip/2}$ is related to $\xi$ via
\EQ{
re^{ip/2}=\frac{1-\xi}{1+\xi}\ .
}
These charges satisfy the relation
\EQ{
J_1 = \sqrt{J_2^2 + 4\sin^2\frac{p}{2}}\>.
\label{Dispersion}
}
In the AdS/CFT context, the components $J_1$ and $J_2$ are identified, up to scaling, with $\Delta-J$ and $Q$, respectively, where $\Delta$ is the scaling dimension of the associated operator in the CFT, and $J$ and $Q$ are two conserved $U(1)$ $R$-charges:
\EQ{
\Delta -J = \frac{\sqrt\lambda}{2\pi}\> J_1\>, \qquad
Q = \frac{\sqrt\lambda}{2\pi}\> J_2\>,
}
where $\lambda$ is the 't~Hooft coupling. Then,~\eqref{Dispersion} becomes the celebrated dispersion relation
\EQ{
\Delta -J = \sqrt{Q^2 + \frac\lambda{\pi^2}\sin^2\frac{p}{2}}\>.
\label{Celebrated}
}
In the rest frame, $\vartheta=0$ or $p=\pi$,
\EQ{
\Delta{\cal Q}_L=2m^{-1}\Lambda\cosec q-2h_\BOmega\cot q
\label{GMcharge}
}
and so
\EQ{
Q=\frac{\sqrt\lambda}\pi\cot q\ .
\label{RCharge}
}

It is important to notice that the internal moduli space of the soliton takes the form of a (co-)adjoint orbit of  $SO(4)$.\footnote{For compact Lie groups there is no distinction between adjoint and co-adjoint orbits.} For the Q-ball kink, this is interpreted as the group of global gauge transformations, while for the giant magnon this is interpreted as the $H=SO(4)$ subgroup of left multiplications of $f$ (modulo gauge transformations) which fix the vacuum $f_0$. In particular the action which fixes the vacuum corresponds to the adjoint action $f\to UfU^{-1}$ which includes a compensating global gauge transformation (right multiplication).\footnote{In the gauge invariant formulation in terms of ${\cal F}=\theta f\theta f^{-1}$, the symmetry action is always vector-like, ${\cal F}\to U{\cal F}U^{-1}$.} The physical interpretation of the moduli space $\mathfrak M$ is that the dyonic motion takes place in a plane perpendicular to the plane picked out by the orbital motion of the vacuum solution. The orientation of this plane is then described by the real Grassmannian $\mathfrak M$ in \eqref{fgt}.

\section{Semi-Classical Quantization of the Dyons}
\label{SCQD}

The soliton that we have constructed is a non-abelian dyon since it carries a  charge under a nonabelian global symmetry group  $H=SO(4)$. The charge is function of the continuous
parameter $q\in(0,\frac\pi2)$ and also of the polarization vector $\BOmega$ that determines the orientation of the motion inside $SO(4)$ and corresponds to the real Grassmannian in \eqref{fgt}. 

Since the soliton is a periodic classical solution, one way to quantize it semi-classically is to use the Bohr-Sommerfeld rule. Remarkably, it leads to different results for the Q-ball kinks and the dyonic giant magnons.
If $\phi$ denotes the soliton solution in its rest frame, then $S[\phi]+MT=2\pi{\EuScript N}$ with ${\EuScript N}=1,2,\ldots$, where  the time period is $T=\pi/(m\cos q)$. 
For the dyonic giant magnon, this gives~\cite{Chen:2006gea}
\EQ{
\text{Dyonic Giant Magnon:}\qquad\qquad\cot q=\frac{\pi{\EuScript N}}{\sqrt\lambda}\ ,\qquad {\EuScript N}=1,2,\ldots,\infty
\label{chr}
}
and so the $U(1)_R$ charge~\eqref{RCharge} is quantized in integer units, $Q={\EuScript N}$, 
which is the known spectrum of the quantized dyonic magnons~\cite{Chen:2006gea}. For the Q-ball kink, the Bohr-Sommerfeld quantization leads to a different quantization of $q$; namely\footnote{The $S^n$ Q-ball kink has an obvious embedding in the $SO(4)/SO(3)$ SSSG theory for which $H=SO(2)$ is abelian. 
In Appendix~\ref{AppB} we show that this yields the complex sine-Gordon theory for which the kink becomes a conventional 
$Q$-ball~\cite{Miramontes:2004dr}. The Bohr-Sommerfeld rule applied to the complex sine-Gordon theory dyon then gives rise to the quantization that follows in~\eqref{iuuCSG}~\cite{Dorey:1994mg}. 
}
\EQ{
\text{Q-Ball Kink:}\qquad\qquad q=\frac{\pi{\EuScript N}}{2k}\ ,\qquad{\EuScript N}=1,2,\ldots,k\ .
\label{iuuCSG}
}
Notice that the dyonic giant magnon spectrum is infinite while the Q-ball kink spectrum is truncated because of the finite range of the parameter $q$. Of course the Bohr-Sommerfeld rule is strictly speaking only valid in the semi-classical regime, so ${\EuScript N}$ of order $\sqrt\lambda$ and $k$, respectively. 

Although the Bohr-Sommerfeld rule gives the energy levels in both cases, it does not reveal the symmetry multiplets at each level. In order to uncover this structure we must proceed in a different way. The idea is to find an effective description of the polarization degree-of-freedom $\BOmega$ which takes values in the moduli space $\mathfrak M$.

In order to motivate the way that we quantize the dyons, it is worth a digression into the quantization of solitons more generally.
Let us consider a hypothetical theory with fields $\phi(x^\mu)$ that has a soliton solution $\phi(X^i;x^\mu)$ having an internal moduli space $\tilde{\mathfrak M}=S^n$ on which the group $SO(n+1)$ acts as a global symmetry of the theory. In the conventional way of proceeding, the existence of a moduli space of solutions, whose coordinates $X^i$ are the collective coordinates of the soliton, means that the equations-of-motion have a set of zero modes
\EQ{
\delta_i\phi=\frac{\partial\phi(X^i;x^\mu)}{\partial X^i}
}
one for each of the collective coordinates. In the conventional setting, for example, for monopoles in Yang-Mills-Higss theories in $3+1$-dimensions, following the philosophy of Manton \cite{Manton:1981mp}, one allows the collective coordinates to depend on time $X^i\to X^i(t)$, and then substitutes $\phi(X^i(t);x^\mu)$ into the action of the theory to extract an effective action for a 1-dimensional theory along the soliton's world line:
\EQ{
S_\text{eff}[X^i]=S[\phi(X^i(t))]=\int dt\,\int dx\,\LAG(\phi(X^i;x^\mu))\ .
}
Assuming that the theory has a quadratic kinetic term, the effective quantum mechanical action that results contains terms which can be linear and quadratic in time derivative:
\EQ{
S_\text{eff}[X^i]=\int dt\, \Big[{\cal Q}_i(X)\dot X^i+\tfrac12g_{ij}(X)\dot X^i\dot X^j + \cdots\Big]\ ,
\label{sole}
}
where 
\EQ{
{\cal Q}_i(X)=\int dx\, \dot\phi\,\delta_i\phi
}
is the charge of the soliton under the symmetry variation $\delta X^i$ and 
$g_{ij}(X)$ is a metric on $\tilde{\mathfrak M}$ given by the inner-product of the zero modes,
\EQ{
g_{ij}(X)=\int dx\,\delta_i\phi\,\delta_j\phi\ .
\label{frr}
}

In many situations the soliton is a static solution and carries no charge. In this case the term linear in $\dot X^i$ is absent and the effective theory is a quantum mechanical sigma model on the moduli space.
To be concrete, and to bring the discussion as close as possible to the present setting, let us suppose that $\tilde{\mathfrak M}=S^n$ with a natural action of $SO(n+1)$ corresponding to some global symmetry of the parent theory.\footnote{Following the analogy, in the present setting, $n=3$ and $S^3$ is the subspace of $S^5$ perpendicular to the plane picked out by the vacuum solution.} The Euler-Lagrange equations that follow from the quantum mechanical sigma model 
\EQ{
S_\text{eff}[X^i]=\int dt\, \tfrac12g_{ij}(X)\dot X^i\dot X^j
\label{sole2}
}
are simply the geodesic equations for the Riemannian manifold $(\tilde{\mathfrak M},g_{ij})$. For the $S^n$ example, the solutions are motions around great circles which carry arbitrary angular momentum, and the motion gives the classical solution with $SO(n+1)$ charge, in other words the excitations are dyonic. 
This system can easily be quantized: the quantum Hamiltonian is the Laplacian on $S^n$ and the states are spherical harmonics. 
This approach is only valid when the correction to the mass is large compared with one-loop quantum corrections but small with respect to the mass of the soliton. This latter requirement is needed in order that the back reaction of the motion on the soliton is small. 

In principle, however, we can include all the effects of the back reaction of the internal motion at the classical level by finding more general dyonic generalizations of the 
original soliton in the parent theory. In the present context it is integrability that allows us to write down the exact dyonic solutions.
The dyonic solutions will have a continuous parameter $\ell$ which is the magnitude of the angular momentum of the motion and also other parameters that determine 
the axis of the rotation, or ``polarization", that is 2 orthonormal vectors $\BOmega_i$. In other words, $\{\BOmega_i\}$ determine the orientation of a plane and therefore the polarization degree-of-freedom takes values in the real Grassmannian
\EQ{
\mathfrak M=\frac{SO(n+1)}{SO(2)\times SO(n-1)}\ .
\label{fgt2}
}
One can also think of this moduli space as a (co-)adjoint orbit of the symmetry group $SO(n+1)$. In fact we can imagine generating the dyon solution from the soliton by performing a time-dependent symmetry transformation, schematically $\phi\to U(t)\phi$ with $U(t)\in SO(n+1)$, and then computing the back-reaction to the motion exactly.
The mass of the dyon will be a function of $\ell$ and the $SO(n+1)$ charge will be of the form 
\EQ{
{\cal Q}=\frac{\ell}2h_\BOmega\ ,\qquad\BOmega=\frac1{\sqrt2}\big(\BOmega_1+i\BOmega_2\big)
\label{Charge}
}
where $h_\BOmega$ is defined as in \eqref{defh}.
 
The effective description of the dyon solution then  takes the form \eqref{sole} but, now, the term linear in $\dot X^i$ is non-vanishing and becomes the dominant term in the semi-classical expansion. We can write the effective description in terms of a time-dependent symmetry transformation of the static solution $U(t)$, in the form~\cite{Hollowood:2010dt}\footnote{There is a slight subtlety here in that the moduli space of the dyon $\mathfrak M$ is not exactly the moduli space of the static solution $\tilde{\mathfrak M}$ because for the time-dependent solutions translations in $t$ act on $\tilde{\mathfrak M}$, and $\mathfrak M$ is the quotient of this action.} 
\EQ{
S_\text{eff}[U]=\int dt\,\Tr\left(U^{-1}\frac{dU}{dt}\,{\cal Q}\right)\ .
\label{cvc2}
}
Below we shall show how to quantize this system. However, it is clear that the equations-of-motion require $U(t)$ to be a constant, and consequently, on shell, the polarization $\BOmega$ is fixed. This is why the quantized dyon does not carry more than one charge and also the collective coordinate dynamics does not contribute to the mass of the dyon since the Hamiltonian vanishes.
In addition, as we show later, the angular momentum must be quantized precisely as $\ell\equiv {\EuScript N}=1,2,\dots$. It turns out that this quantization rule is identical to the Bohr-Sommerfeld quantization of the dyon; however, the bonus of this method is that we can compute the multiplet structure of the levels. 
We shall find that the states of a given ${\EuScript N}$ transform in the rank-$\EuScript N$ symmetric representation of $SO(n+1)$. The mass of the dyon is then just the classical mass  but with the quantized values of $\ell$ inserted. 

The semi-classical quantization of the dyonic giant magnon and Q-ball kink fits exactly into the story above apart from the fact that in the case of the SSSG theory there is no original uncharged soliton solution to start with. The parameter $q$ determines the charge of the soliton  as in \eqref{Charge}.
For the Q-ball kink, it plays the r\^ole of $\ell$ as in \eqref{nxn} below. Then, as $q\to0$,
the mass of the kink 
goes to 0 as $\ell\to0$ and so solitons with small charge are not bona-fide semi-classical objects; rather they are actually the perturbative excitations of the theory~\cite{Hollowood:2010dt}. For the dyonic giant magnon the situation is more complicated due to the fact that the system has no relativistic invariance. In this case, as $\ell$ goes to zero, with $\ell$ as in \eqref{nzn} below, the dyonic magnon becomes an ordinary giant magnon which is a semi-classical object when $\sin\frac p2$ is of order 1 (in~\eqref{Celebrated}, $Q$ is small but $\Delta -J$ is large), but becomes a perturbative excitation as $p\to0$ (both $Q$ and $\Delta -J$ are small). In both situations, however,
the direct quantization of the dyonic solution as described above is valid for large enough $\ell$.

The claim is that when we take the dyon solution and transform it with a time-dependent $SO(4)$ transformation, substitute it into the action and perform the spatial integral we obtain an effective action of the form \eqref{cvc2}. For the Q-ball kinks of the SSSG theory this was shown in \cite{Hollowood:2010dt} with the result
\EQ{
\text{Q-Ball Kink:}\qquad\qquad \ell(q)=\frac{2kq}\pi\ .
\label{nxn}
}
Here, $k^{-1}$ plays the r\^ole of the coupling.
Now we show that we get a similar action for the dyonic giant magnon. According to~\eqref{Match}, in the  $A_\mu=0$ on-shell gauge we have $B_-=0$ and $B_+=\gamma^{-1}\partial_+\gamma$, and we take
\EQ{
f\longrightarrow U(t)fU(t)^{-1}\ ,\qquad U(t)\in SO(4)\ .
}
So for 
\EQ{
J_+=f^{-1}\partial_+f-B_+=f^{-1}\partial_+f-\gamma^{-1}\partial_+\gamma\ ,
}
and given that $\gamma\to U(t)\gamma U(t)^{-1}$ also,
we obtain
\EQ{
J_+ &\longrightarrow U\Big(J_+ + f^{-1}[U^{-1}\dot U,f]-\gamma^{-1}[U^{-1}\dot U,\gamma]\Big)U^{-1}\ ,\\
J_- &\longrightarrow U\Big(J_- + f^{-1}[U^{-1}\dot U,f]\Big)U^{-1}\ .
}
Then, to leading order in the semi-classical approximation, we can work to linear order in $\dot U$, 
\EQ{
\delta S&=-\frac{\sqrt\lambda}{4\pi}\int d^2x\,\Tr\big(\delta J_+J_-+J_+\delta J_-\big)\\
&=-\frac{\sqrt\lambda}{4\pi}\int d^2x\,\Tr\Big[(f^{-1}U^{-1}\dot Uf-\gamma^{-1}U^{-1}\dot U\gamma)J_-\\
&\qquad\qquad+(f^{-1}U^{-1}\dot Uf-U^{-1}\dot U)J_+\Big]\\
&=-\frac{\sqrt\lambda}{4\pi}\int_{-\infty}^\infty dt\,\Tr\Big(U^{-1}\dot U
\,\Delta{\cal Q}_L\Big)
}
where, using the Virasoro constraints \eqref{virc} and  $fJ_\mu f^{-1}=D_\mu f\,f^{-1}$, we have identified
\EQ{
\int_{-\infty}^\infty dx\, \big(
D_0f\,f^{-1}-D_0f_0\,f_0^{-1}\big)=\Delta{\cal Q}_L\ .
}
Therefore, just like the SSSG case, we have an effective quantum mechanical action
\EQ{
S_\text{eff}[U]=\frac{\ell(q)}2\int_{-\infty}^\infty dt\,\Tr\Big(U^{-1}\frac{dU}{dt}
\,h_\BOmega\Big)\ .
}
where, using~\eqref{GMcharge}, 
\EQ{
\text{Dyonic Giant Magnon:}\qquad\qquad\ell(q)=\frac{\sqrt\lambda}{\pi}\cot q\ .
\label{nzn}
}

We now proceed to a quantization of an action of the form \eqref{cvc2}. Before proceeding in earnest, it is worthwhile making some comments about such theories. Unlike the soliton effective theory \eqref{sole}, the action in \eqref{cvc2} does not involve metric data of the moduli space $\mathfrak M$. In fact it lies in the class of ``topological" or Chern-Simons quantum mechanics defined and investigated in \cite{Dunne:1989hv,Howe:1989uk,Ivanov:2003qq}. Such a system is defined by a manifold $\mathfrak M$ with a symplectic 2-form $F$ which can be written locally as $F=dA$. The quantum mechanical system has the form
\EQ{
S=\frac \ell2 \int f^*A\ ,
}
where $f^*A$ is the pull-back of $A$ to a one cycle (the world-line in our case). The normalization of $F$ is determined by $\int _{\mathfrak M}F\wedge\cdots\wedge F=4\pi$ and $\ell$ is a coupling constant. For the SSSG theories all the spaces $\mathfrak M$ are actually homogeneous K\"ahler manifolds, in which case the sympletic form can be taken to be the K\"ahler form and the action can be written in terms of the K\"ahler potential $K$,
\EQ{
S=\frac \ell2\int dt\,\Big(-i\dot z^i\frac{\partial K}{\partial z^i}+\text{c.c.}\Big)\ .
}
where $(z^i,\bar z^i)$ are a set of complex coordinates for $\mathfrak M$. A rather beautiful way to quantize the theory is to use ``analytic quantization" as described in \cite{Ivanov:2003qq}. Wavefunctions are sections of holormorphic line bundles over $\mathfrak M$ with curvature $\ell F/(4\pi)$ which must be integral for consistency. Given the normalization of $F$ this requires that $\ell$ is quantized in integer units. Below we shall explain how to quantize  the particular example \eqref{fgt}, or the more general \eqref{fgt2}, which are homogeneous K\"ahler manifolds, in a more pedestrian 
way.

First of all, we can think of the time-dependence via the vector $\BOmega$, by identifying $\BOmega(t)=U(t)\BOmega_0$, where $\BOmega_0$ is some fixed reference vector. 
In this case, the effective action takes the form
\EQ{
S_\text{eff}=-\frac{i\ell}2\int dt\,\Big(\BOmega^*\cdot\frac{d\BOmega}{dt}-\BOmega\cdot\frac{d\BOmega^*}{dt}\Big)=-i\ell\int dt\,\BOmega^*\cdot\frac{d\BOmega}{dt}
}
and implicitly we have the constraints $\BOmega^*\cdot\BOmega=1$ and $\BOmega\cdot\BOmega=0$ as well as the identification $\BOmega\sim e^{i\alpha}\BOmega$.
It is useful to relax the constraints and
enlarge the phase space to ${\mathbb C}^4$ since then the Poisson brackets are trivial: 
\EQ{
\{\BOmega_i,\BOmega^*_j\}=\frac{i}{\ell}\delta_{ij}\ .
} 
The way to reduce the larger phase space proceeds via a K\"ahler quotient. This starts by noticing that 
the $U(1)$ symmetry $\BOmega\to e^{i\alpha}\BOmega$ 
is a Hamiltonian symmetry generated by 
$\Phi=\BOmega^*\cdot\BOmega$. The physical phase space corresponds to restricting ${\mathbb C}^4$ to the level set 
\EQ{
\Phi=\BOmega^*\cdot\BOmega=1
}
and performing a quotient by the $U(1)$ symmetry, as well as imposing the constraint $\BOmega\cdot\BOmega=0$.

In the quantum theory, we can replace the Poisson brackets by commutators 
involving the operators $\hat\BOmega_i$ and $\hat\BOmega_i^\dagger$:
\EQ{
[\hat\BOmega_i,\hat\BOmega_j^\dagger]=\frac{1}{\ell}\delta_{ij}
\label{nco}
}
and build a Hilbert space by treating the former as annihilation operators and the latter as creation operators. The generator of the Hamiltonian symmetry 
\EQ{
\hat\Phi=\hat\BOmega^\dagger\cdot\hat\BOmega =\frac{\hat{\EuScript N}}\ell
}
is proportional to the number operator $\hat{\EuScript N}$ and the constraint $\hat\Phi=1$, along with the quantization of the occupation number, implies the quantization of $\ell$:
\EQ{
\ell={\EuScript N}=1,2,\ldots\ .
\label{iuu}
}
The Hilbert space is spanned by the states\footnote{Notice that the quotient by $U(1)$ is trivial at the level of the Hilbert space.}
\EQ{
\hat\BOmega_{i_1}^\dagger\hat\BOmega_{i_2}^\dagger\cdots\hat\BOmega_{i_{\EuScript N}}^\dagger|0\rangle\ ,
}
However, there is the additional constraint $\BOmega\cdot\BOmega=0$ to impose. Since this is holomorphic
it can be implemented directly at the level of the Fock space by removing by hand states which are of the form $\sum_i\cdots\hat\BOmega_i^\dagger\cdots\hat\BOmega_i^\dagger\cdots |0\rangle$.  One recognizes this as the process of ``removing traces" that is well-known in the Young Tableaux approach to the orthogonal groups. The remaining states form a representation space for the rank-${\EuScript N}$ symmetric representations of $SO(4)$. 

The construction we have presented has an interesting interpretation as ``fuzzy geometry''~\cite{Balachandran:2005ew}. This follows from the fact that  in the quantum theory the coordinates of $\mathfrak M$ do not commute as in \eqref{nco}.
However, as ${\EuScript N}$ increases the non-commutativity 
gets less marked and the fuzzy geometry 
becomes a closer approximation of the classical geometry in the limit ${\EuScript N}\to\infty$, which clearly requires $k\to\infty$, the semi-classical limit.  

One can see now that the quantization of $q$ in both cases is equivalent to the Bohr-Sommerfeld rule in \eqref{chr} and \eqref{iuuCSG}.
In both cases, the states come in the ${\EuScript N}$-rank symmetric representations of $SO(4)$, but the giant magnon tower is unbounded whereas the Q-ball kink tower is bounded to have height $k$. In addition, for the giant magnons, the quantization of ${\EuScript N}$ is the expected quantization of the $R$ charge~\cite{Chen:2006gea}. 
Correspondingly, as a result of the quantization of $q$, the continuous spectrum of classical Q-ball kinks becomes discrete,
\EQ{
M=\frac{4km}\pi\sin\Big(\frac{\pi{\EuScript N}}{2k}\Big)\ ,\qquad 
{\EuScript N}=1,2,\ldots,k\ .
}

Strictly speaking the semi-classical analysis only applies when ${\EuScript N}$ is of order $k$ and $\sqrt\lambda$ for the two cases, respectively. However, the results appear to apply also for small $\EuScript N$. For the SSSG theory, the exact S-matrix constructed for the case $F/G=\CP^n$ in~\cite{Hollowood:2010rv} suggests that in the full quantum theory there is simply a finite renormalization of $k$, which is a well-known feature of the WZW theory.
The states at the bottom of the tower ${\EuScript N}=1$, which transform in the vector representation of $SO(4)$, correspond to the perturbative excitations of the SSSG Lagrangian~\cite{Hollowood:2010dt}. In particular, the gapless excitations in the classical theory get a mass gap at the quantum level.

\vspace{0.5cm}

\acknowledgments

TJH would like to acknowledge the support of STFC grant
ST/G000506/1.

\noindent 
JLM acknowledges the support of MICINN 
(grants FPA2008-01838 and\break 
FPA2008-01177), Xunta de Galicia (Consejer\'\i a de Educaci\'on and INCITE09.296.035PR), the 
Spanish Consolider-Ingenio 2010
Programme CPAN (CSD2007-00042), and FEDER.

\noindent
We would both like to
thank Arkady Tseytlin and Ben Hoare for discussions and comments on an earlier draft.

\startappendix

\Appendix{A Tale of Two Dressings}
\label{AppA}

In this appendix, we relate the sigma model dressing formalism, as described in \cite{Hollowood:2009tw},\footnote{In order to compare with that reference one must re-scale $\Lambda\to2\Lambda$.} with the SSSG dressing formalism described in detail in \cite{Hollowood:2010dt} and summarized in Section~\ref{Dressing}. Quantities in the sigma model dressing formalism, if denoted with the same letter as in the SSSG formalism, will be highlighted with a tilde.

In the sigma model formalism in terms of the gauge invariant field ${\cal F}=\theta f\theta f^{-1}$, the central quantity is $\Psi(\lambda)$, which is the solution of the linear system
\EQ{
\Big[\partial_\pm-\frac{\partial_\pm{\cal F}\,{\cal F}^{-1}}{1\pm\lambda}\Big]
\Psi(\lambda)=0\ ,
}
(the spectral parameter $\lambda$ here is not to be confused with the 't~Hooft coupling).
The dressing method starts with
\EQ{
\Psi(\lambda)=\tilde\chi(\lambda)\Psi_0(\lambda)\ ,
}
where
\EQ{
\Psi_0(\lambda)=\exp\Big[\frac{2x^+}{1+\lambda}\Lambda+\frac{2x^-}{1-\lambda}
\Lambda\Big]
}
is the ``vacuum" solution. Then, the dressing transformation takes the form
\EQ{
\tilde\chi(\lambda)=1+
\frac{\tilde\BF_j\tilde\Gamma^{-1}_{ji}\tilde\BF_i^\dagger}{\lambda-\lambda_i}\ .
}
Here, $\tilde\BF_i$ are complex 6-vectors given by
\EQ{
\tilde\BF_i=\Psi_0(\lambda_i^*)\Bvarpi_i\ ,
}
where the $\Bvarpi_i$ are the same constant vector as in the SSSG formalism \eqref{vwr},
\EQ{
\tilde\Gamma_{ij}=\frac{\tilde\BF_i^*\cdot \tilde\BF_j}{\lambda_i-\lambda^*_j}\ ,
}
and the poles are related to those in~\eqref{Poles} by means of~\eqref{RelPole}.

The gauge invariant field is given by
\EQ{
{\cal F}=\Psi(0)\ ,
}
and the SSSG field $\gamma$ is given in the two formalisms by \cite{Hollowood:2009tw}
\EQ{
\gamma=\chi(0)^{-1}={\cal F}_0^{-1/2}\tilde\chi^{-1}(1)\tilde\chi(-1){\cal F}_0^{1/2}\ ,
\label{Gamma}
}
where ${\cal F}_0=\Psi_0(0)$ is the vacuum solution. One finds that the two formalisms are simply related via
\EQ{
\Upsilon_0(z)={\cal F}_0^{-1/2}\Psi_0(\lambda)\ ,
}
which requires
\EQ{
z=\frac{1-\lambda}{1+\lambda}\ .
}
This means that
\EQ{
\BF_i={\cal F}_0^{-1/2}\tilde\BF_i\ ,\qquad 
\xi_i=\frac{1-\lambda_i}{1+\lambda_i}\ .
\label{RelPole}
}
Furthermore, one can show that
\EQ{
-\frac1{\xi_j}\Gamma^{-1}_{ij}=\frac2{(1-\lambda_j)(1+\lambda_i^*)}\tilde\Gamma^{-1}_{ij}\ ,
}
from which it follows
\EQ{
\gamma^{-1}=\chi(0)=1-\frac{\BF_i\Gamma^{-1}_{ij}\BF^\dagger_j}{\xi_j}&={\cal F}_0^{-1/2}\Big(1+\frac{2\tilde\BF_i\tilde\Gamma^{-1}_{ij}\BF_j^\dagger}
{(1-\lambda_j)(1+\lambda_i^*)}\Big){\cal F}_0^{1/2} \\[5pt]
&={\cal F}_0^{-1/2}\tilde\chi^{-1}(-1)\tilde\chi(1){\cal F}_0^{1/2}\ ,
}
which reproduces~\eqref{Gamma}.

The current associated to the left action $f\to Uf$, with $U\in H$, is $D_\mu f\, f^{-1}$. 
Using the equations of motion of $B_\mu$, one can easily show that
\EQ{
D_\mu f\,f^{-1}=\frac12\theta\partial_\mu{\cal F}\,{\cal F}^{-1}\theta\ .
}
Moreover, in \cite{Hollowood:2009tw} it was shown that
\EQ{
\int_{-\infty}^\infty dx\,\big(\partial_0{\cal F}\,{\cal F}^{-1}-\partial_0{\cal F}_0\,
{\cal F}_0^{-1}\big)
=\tilde F_i\tilde\Gamma^{-1}_{ij}\tilde F_j^\dagger\Big|_{x=\infty}-
\tilde F_i\tilde\Gamma^{-1}_{ij}\tilde F_j^\dagger\Big|_{x=-\infty}
}
and from this one finds~\cite{Hollowood:2009sc}
\EQ{
\Delta{\cal Q}_L=-\frac i2\big(\lambda-\lambda^{-1}-\lambda^*+\lambda^{*-1}\big)
m^{-1}\Lambda+\frac i2\big(\lambda+\lambda^{-1}-\lambda^*-\lambda^{*-1}\big)h_\BOmega\ ,
}
where $\lambda=(1-\xi)/(1+\xi)= r e^{ip/2}$.
 
\Appendix{$S^n$ Kinks as Complex Sine-Gordon Q-balls}
\label{AppB}

The $F/G=S^n$ Q-ball kinks have an obvious embedding in $SO(4)/SO(3)$. This yields the complex sine-Gordon theory where the kinks become conventional Q-balls~\cite{Miramontes:2004dr,Bowcock:2008dn}. In this appendix we shall write the SSSG action corresponding to a Q-ball kink in terms of the Lagrangian of the complex sine-Gordon theory 
\SP{
{\cal L}_\text{CSG}[\psi,\lambda]= \frac{\partial_\mu\psi \partial^\mu\psi^\ast}{1-\psi\psi^\ast} -\lambda \psi\psi^\ast\,.
\label{CSGlag}
}
The resulting expression provides the identification of the Bohr-Sommerfeld quatization of kinks with the quantization of Q-balls in the complex sine-Gordon theory~\cite{Dorey:1994mg} that leads to~\eqref{iuuCSG}. Remarkably, the result makes use of the boundary term~\eqref{Topol} in a rather non-trivial way.

For $\gamma\in SO(3)\subset SO(4)$ and $A_\pm \in{\mathfrak so}(3)$, we will use the following parameterization of Euler-angle type:
\EQ{
\gamma = e^{(\alpha+\beta) r_1} \,e^{\theta r_3}\, e^{(\alpha-\beta) r_1}\ ,\qquad
A_\pm = a_\pm r_1\,,
\label{Param}
}
where
\EQ{
r_1= \MAT{0&0&0&0\\0&0&0&0\\0&0&0&1\\0&0&-1&0}=\Be_3 \Be_4^t -\Be_4 \Be_3^t\,,\qquad
r_3= \MAT{0&0&0&0\\0&0&1&0\\0&-1&0&0\\0&0&0&0}\Be_2 \Be_3^t -\Be_3 \Be_2^t\ ,
}
and $\Be_a$ denotes the $n+1$~column vector with components $(\Be_a)_i=\delta_{i,a}$.
Then, a particular embedding of the $SO(4)/SO(3)$ Q-ball kink into $S^n=SO(n+1)/SO(n)$ is specified by the polarization vector $\BOmega=\frac{1}{\sqrt{2}}(\BOmega_1+i\BOmega_2)$ by means of
\SP{
r_1  \longrightarrow \BOmega_1 \BOmega_2^t - \BOmega_2 \BOmega_1^t =h_\BOmega\,,\qquad
r_3 \longrightarrow \Be_2 \BOmega_1^t - \BOmega_1 \Be_2^t 
\,.
}
For $SO(4)/SO(3)$, the SSSG action~\eqref{ala} is invariant under abelian $H=SO(2)$ vector gauge transformations, which correspond to
\EQ{
\beta \to \beta +\rho\,,\qquad a_\pm \to a_\pm-\partial_\pm \rho\,,
}
while $\theta$ and $\alpha$ remain invariant. These transformations also leave invariant the combinations of fields
\EQ{
b_\pm = a_\pm + \partial_\pm \beta\,.
}
Then,
\SP{
S_\text{gWZW}=&\frac k{4\pi}\int d^2x\, \Big[2\partial_+\theta \partial_-\theta
+ 8 \cos^2(\theta/2)\Big( \partial_+\alpha \partial_-\alpha
+b_+\partial_-\alpha - b_-\partial_+\alpha\Big) \\[5pt]
&\qquad+8\sin^2(\theta/2)\,b_+b_-  
+8\cos^2(\theta/2) \big(\partial_+\alpha \partial_-\beta - \partial_+\beta \partial_-\alpha\big) \Big]\\[5pt]
-&\frac k{\pi}\int d^3x\,\epsilon^{abc} \,\partial_a\big[\cos^2(\theta/2) \,\partial_b\alpha \,\partial_c\beta\Big]\,.
\label{Action0}
}
The last contribution corresponds to the WZ term, whose form has to be determined by the conditon of gauge invariance. In this case, it can be partially fixed by removing all the terms that depend explicitly on $\beta$. The result is
\SP{
S_\text{gWZW}=&\frac k{4\pi}\int d^2x\, \Big[2\partial_+\theta \partial_-\theta 
+8 \cot^2(\theta/2)\partial_+\alpha \partial_-\alpha
\\[5pt]
&
\hspace{-0.5cm}
+8\sin^2(\theta/2)\big(b_+ -\cot^2(\theta/2)\partial_+\alpha\big)\, 
\big(b_- +\cot^2(\theta/2)\partial_-\alpha\big)
+ \epsilon^{\mu\nu} \partial_\mu F_\nu\Big]\,,
\label{Action1}
}
where $F_\mu$ parameterizes the remaining ambiguities coming from the WZ term. 

If we ignore the boundary term~\eqref{Topol}, we can take $F_\mu=0$. Then, using the equations of motion of $a_\pm$, 
\SP{
b_\pm = \pm \cot^2(\theta/2)\partial_\pm\alpha\,,
}
the SSSG action becomes
\SP{
&S_\text{gWZW}[\gamma,A_\mu] -\frac k{2\pi}\int d^2x\,\Tr\left(\Lambda
\gamma^{-1}\Lambda\gamma-\Lambda^2\right)\\[5pt]
&\qquad\qquad
=\frac k{2\pi}\int d^2x\,\Big({\cal L}_\text{CSG}[\cos(\theta/2)e^{i\alpha},-4m^2] - 4m^2\Big)\,,
\label{CSGneg}
}
which involves the complex sine-Gordon Lagrangian with negative mass term. This Lagrangian has a $U(1)$ degenerate set of vacua with $|\psi|=1$. At rest, its soliton solutions are time independent (non-dyonic) kinks that interpolate between two different vacua~\cite{Lund:1976ze,Lund:1977dt} (see also~\cite{Miramontes:2004dr}). In addition, notice that this Lagrangian does not
have a good expansion in terms of fields around their vacuum values due to the $\cot^2 \theta$ term
in~\eqref{Action1}, or $(1-|\psi|^2)^{-1}$ in~\eqref{CSGlag}.

However, the full SSSG action does include the boundary term~\eqref{Topol}. Then, the condition of gauge invariance fixes\footnote{Notice that $\Tr\big(r_1\,\phi\big)$ is gauge invariant. Moreover, using the Baker-Campbell-Hausdorff formula,
\SP{
\Tr\big(r_1\,\phi\big)= -4\alpha + \frac{1}{3} \alpha\, \theta^2+\cdots\,, 
\label{BCH}
}
and all the terms in the ellipsis are proportional to $\theta^2$.
}
\SP{
F_\mu = - \Tr\big(r_1\,\phi\big)\,\partial_\mu\beta,
}
and the true SSSG action~\eqref{ala} reads
\SP{
S=&\frac k{4\pi}\int d^2x\, \Big[2\partial_+\theta \partial_-\theta 
+8 \cot^2(\theta/2)\partial_+\alpha \partial_-\alpha
-8m^2 \sin^2(\theta/2)\\[5pt]
&
\hspace{-0.5cm}
+8\sin^2(\theta/2)\big(b_+ -\cot^2(\theta/2)\partial_+\alpha\big)\, 
\big(b_- +\cot^2(\theta/2)\partial_-\alpha\big)
- \epsilon^{\mu\nu} \partial_\mu\big(b_\nu\, \Tr(r_1\,\phi)\big)\Big]\,.
\label{Action2}
}
Now, we can use the equations of motion of $a_\pm$ and $\alpha$,
\EQ{
b_\pm = \pm \cot^2(\theta/2)\partial_\pm\alpha\,,\qquad
\partial_+ b_- -\partial_-b_+=0\,,
\label{eom1}
}
to write $b_\pm$ in terms of a new field $\varphi$ as follows
\EQ{
b_\pm = \pm \cot^2(\theta/2)\partial_\pm\alpha = \partial_\pm\varphi\,.
\label{eom2}
}
This provides an explicit relation between the SSSG action for the $S^n$ kink Q-ball and the complex sine-Gordon Lagrangian~\eqref{CSGlag} with positive mass term
\SP{
S =
\frac k{2\pi}\int d^2x\, \Big({\cal L}_\text{CSG}[\sin(\theta/2)e^{i\varphi},+4m^2] -\frac{1}{2}  \epsilon^{\mu\nu}\partial_\mu\big[\big(\Tr(r_1\,\phi)+4\alpha\big)\, \partial_\nu\varphi\big]\Big)\,.
\label{CSGpos}
}
This Lagrangian has a non-degenerate vacuum at $|\psi|=0$, and its soliton solutions are $Q$-balls that carry $U(1)$ Noether charge~\cite{Getmanov:1977hk} (see also~\cite{Miramontes:2004dr}). For these solutions at rest, $\varphi$ only depends on $t$ and $\Tr(r_1\,\phi)+4\alpha$ vanishes at $x=\pm\infty$. Therefore, the last term in~\eqref{CSGpos} vanishes and the SSSG action for a kink in $S^n$ is equal to the action of a complex sine-Gordon $Q$-ball, which is the result used in Section~\ref{SCQD} to quantize the $S^n$ SSSG kinks by means of the Bohr-Sommerfeld rule, eq.~\eqref{iuuCSG}.

Finally, we can use the particular example of the $SO(4)/SO(3)$ SSSG theory to illustrate the need to supplement the SSSG action with the boundary term~\eqref{Topol}. Consider the action~\eqref{Action2} for $\gamma\in H=SO(2)$, which corresponds to $\theta=0$,
and leave the normalization of the boundary term free.\footnote{Notice that $\beta$ is not a good coordinate around $\theta=0$, in the same way that the polar angle is not a good coordinate around $r=0$.}  Then, using~\eqref{Action0} and~\eqref{BCH},
\SP{
S_\text{gWZW} &+ NS_\text{bt}=\frac {2k}{\pi}\int d^2x\, \Big[\partial_+\alpha \partial_-\alpha
+a_+\partial_-\alpha - a_-\partial_+\alpha\\[5pt]
&
+N\big(\partial_+(a_-\,\alpha)- \partial_-(a_+\,\alpha)\big)\Big] -\frac k{\pi}\int d^3x\,\epsilon^{abc} \,\partial_a\big[\partial_b\alpha \,\partial_c\beta\Big]\,,
}
which shows that the naive WZ term vanishes.
Then, the choice $N=1$ is singled out as the only one that ensures gauge invariance:
\SP{
S_\text{gWZW} + S_\text{bt}=&\frac {2k}{\pi}\int d^2x\, \Big[\partial_+\alpha \partial_-\alpha
+\big(\partial_+ a_--\partial_- a_+\big)\alpha\Big]\,. \\[5pt]
}

\end{document}